\documentclass[prl,twocolumn,groupedaddress,superscriptaddress,showpacs,floatfix]{revtex4-1}
\usepackage[dvips]{graphicx}
\usepackage{dcolumn}
\usepackage{bm}
\usepackage{amsmath}
\usepackage{amssymb}
\usepackage{revsymb4-1}
\usepackage{hhline}

\begin{document}

\title{Hyperbolic Metamaterials with Bragg Polaritons}
\author{Evgeny S. Sedov}
\affiliation{Department of Physics and Applied Mathematics, Vladimir State
University named after A. G. and N. G. Stoletovs, Gorky str. 87, 600000,
Vladimir, Russia}
\affiliation{National Research University for Information Technology,
Mechanics and Optics (ITMO), St. Petersburg 197101, Russia}
\author{I. V. Iorsh}
\affiliation{National Research University for Information Technology,
Mechanics and Optics (ITMO), St. Petersburg 197101, Russia}
\author{S. M. Arakelian}
\affiliation{Department of Physics and Applied Mathematics, Vladimir State
University named after A. G. and N. G. Stoletovs, Gorky str. 87, 600000,
Vladimir, Russia}
\author{A. P. Alodjants}
\affiliation{Department of Physics and Applied Mathematics, Vladimir State
University named after A. G. and N. G. Stoletovs, Gorky str. 87, 600000,
Vladimir, Russia}
\affiliation{Russian Quantum Center, Novaya 100, 143025
Skolkovo, Moscow Region, Russia}
\email[Electronic address: ]{alodjants@vlsu.ru}
\author{Alexey Kavokin}
\affiliation{Russian Quantum Center, Novaya 100, 143025
Skolkovo, Moscow Region, Russia}
\affiliation{Spin Optics Laboratory, St. Petersburg State
University, Ul'anovskaya, Peterhof, St. Petersburg 198504, Russia}
\affiliation{School of Physics and Astronomy, University of Southampton,
SO17 1NJ Southampton, United Kingdom}

\begin{abstract}
We propose a novel mechanism for designing quantum hyperbolic metamaterials
with use of semiconductor Bragg mirrors containing periodically arranged
quantum wells. The hyperbolic dispersion of exciton-polariton modes is
realized near the top of the first allowed photonic miniband in such
structure which leads to formation of exciton-polariton X-waves.
Exciton-light coupling provides a resonant non-linearity which leads to
non-trivial topologic solutions. We predict formation of low amplitude
spatially localized oscillatory structures: oscillons described by kink
shaped solutions of the effective Ginzburg-Landau-Higgs equation. The
oscillons have direct analogies in the gravitational theory. We discuss
implementation of exciton-polariton Higgs fields for the Schr\"{o}dinger cat
state generation.
\end{abstract}

\maketitle

\emph{Introduction} A remarkable similarity between propagation of
electromagnetic waves in inhomogeneous media described by the Maxwell's
equations and propagation of photons in curved space-time described by the
general relativity laws offers a possibility of designing media where light
propagates along pre-defined curved trajectories. This concept, known as
\emph{transformation optics}~\cite{NatureMaterials93872010} not only allowed
emulating many gravitational effects such as gravitational lensing~\cite%
{NatPhoton79022013}, event horizon~\cite{ProgressinOptics53692009} etc, but
also led to a bunch of intriguing practical applications, such as optical
cloaking~\cite{ContemporaryPhysics542732013}, and superresolution optical
imaging~\cite{PRL8539662000}. Construction of the media with predefined
profiles of electric and magnetic permeabilities is feasible with use of
\emph{metamaterials}~\cite{IEEETrans4720751999}, artificial periodic
structures whose optical properties are governed both by the electromagnetic
response of individual structure elements and by the geometry of the
lattice. Hyperbolic metamatrials (HMM's) are highly anisotropic media that
have hyperbolic (or indefinite) dispersion~\cite{NatPhoton79582013},
determined by their effective electric and/or magnetic tensors. Such
structures represent the ultra-anisotropic limit of traditional uniaxial
crystals. One of the diagonal components of either permittivity ($%
\varepsilon $) or permeability ($\mu $) tensors of HMM has an opposite sign
with respect to the other two diagonal components.
\begin{figure}[tbp]
\includegraphics[width= 0.95\columnwidth]{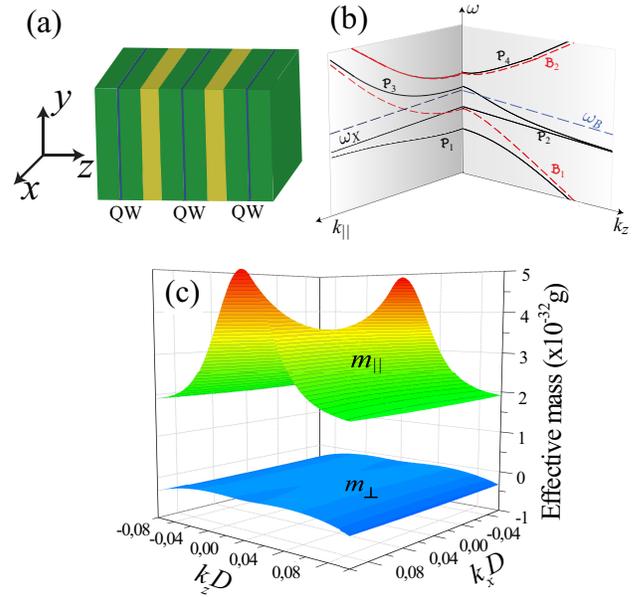}
\caption{(a) -- Schematic picture of spatially periodical structure
(\textquotedblleft Bragg mirror\textquotedblright ), (b) -- dispersion
characteristics for four Bragg exciton-polariton branches (black solid
lines); photonic Bragg mode dispersion is shown with red dashed lines, and,
(c) is the effective mass dispersion for the lower polariton branch $%
\mathcal{P}_{1}$. }
\label{FIGschemea}
\end{figure}
Recently, HMMs have attracted an enhanced attention both due to the
promising applications in quantum lifetime engineering~\cite%
{ApplPhysB1002152010} and subwavelength image transfer~\cite%
{Science31516862007}, and because of their relatively low production costs
as compared to other optical metamaterial designs. In contrast to
all-dielectric uni-axial anisotropic media for which both dielectric
permittivities are positive: $\varepsilon _{1}\equiv \varepsilon
_{x}=\varepsilon _{y}>0$ and $\varepsilon _{2}\equiv \varepsilon _{z}>0$, in
HMMs $\varepsilon _{1}$ and $\varepsilon _{2}$ have opposite signs in some
frequency range due to the presence of metallic layers. As a result, the
analogy between wave propagation governed by the Helmholtz equation and the
effective Klein-Gordon equation for massive particle with fictitious time
coordinate can be obtained for the description of a coherent CW laser beam
propagation in such a structure. This analogy makes possible creation of the
Minkowski space-time using HMMs~\cite{PRA880338432013}. Applications of HMMs
for modelling gravity and cosmology problems within scalar $\phi ^{4}$-field
theories require introducing a strong Kerr-like nonlinearity to the medium, ~%
\cite{RubakovClassicalTheoryofGaugeFields2002}. However, the nonlinear
response of the conventional HMMs is relatively weak. Moreover, most of the
studied HMMs are essentially periodic arrays of metallic inclusions,
characterised by large ohmic losses and decay of the electromagnetic field
propagation.

To overcome these problems, in this Letter we propose a novel approach for
emulating quantum effects in curved space-time using resonant semiconductor Bragg mirrors.
Light propagation in such structures has been extensively studied both in theory and in experiments~\cite{FizTverdTela3621181994,JApplPhys795951996,PRA640138092001,PRB80121306R2009, KavokinBaumbergMicrocavities2007,PRL1060764012011}.
In particular, the dramatic modulation of the reflectivity spectra of Bragg spectra in the vicinity of exciton resonances has been predicted in~\cite{JApplPhys795951996}.
The  peculiar dispersion of mixed photon-exciton modes in Bragg arranged quantum wells (QWs) has been discussed in literature (see, e.g.~\cite{PRA640138092001,PRB80121306R2009,KavokinBaumbergMicrocavities2007,PRL1060764012011}).
Here we show that planar periodic semiconductor Bragg mirror structures with embedded QWs allow for controlling the signs of effective masses of mixed light-matter quasiparticles termed \emph{Bragg exciton-polaritons} in order to create a quantum HMM.
Exciton-polaritons are responsible for the strong nonlinear dielectric susceptibility of the system due to their excitonic part, while their dispersion properties can be tailored through the photonic part by tuning the layer thicknesses~\cite{PRB80121306R2009}.
The magnitude and sign of the polariton effective mass in such structures affect the effective dielectric permeabilities which are crucial for designing HMMs, cf.~\cite{PRA890636242014}.

\emph{Bragg mirror model.} The semiconductor structure that we are discussing here is schematically shown in Fig.~\ref{FIGschemea}(a).
It consists of the periodic array of alternating dielectric layers with QWs placed in the centres of the layers of one type.
The exciton frequency is tuned to the high frequency edge of the second photonic band gap, which is characterized by the saddle point in the dispersion surface, cf.~\cite{PRB80121306R2009}.
The Hamiltonian of the structure in Fig.~\ref{FIGschemea} has a generic structure, that is $\hat{H}=\hat{H}_{%
\mathrm{ph}}+\hat{H}_{X}+\hat{H}_{\mathrm{coup}}+\hat{H}_{\mathrm{nl}}$
where $\hat{H}_{\mathrm{ph}}$ is the photonic part, $\hat{H}_{X}$ is the
excitonic part, $\hat{H}_{\mathrm{coup}}$ accounts for the exciton-photon
coupling, and $\hat{H}_{\mathrm{nl}}$ is the nonlinear part, originated from
the exciton-exciton scattering. If we consider the frequencies in the
vicinity of the second photonic band gap of the Bragg mirror, we can
diagonalize the linear part of the Hamiltonian $\hat{H}$, using the approach
described in~\cite{PRA640138092001} to obtain the dispersions of the four
polariton branches $\mathcal{P}_{1}$, $\mathcal{P}_{2}$, $\mathcal{P}_{3}$, $%
\mathcal{P}_{4}$, which are shown in Fig.~\ref{FIGschemea}(b);
for more details see the supplemental material~\cite{SuM}.

The lower branch (LB) of polaritons is characterized by the effective mass tensor whose diagonal components differ in sign.
In general the effective mass tensor is dispersive due to the non-parabolicity of the polariton band, however for the wave vectors small compared to the inverse period of the structure $1/D$, one can safely assume the tensor components constant (see Fig.~\ref{FIGschemea}(c)).
Note that due to the smallness of the effective mass, exciton polaritons remain within the light cone, i.~e. at the wave vectors smaller than $1/D$, even at room temperature.

In general case, the Hamiltonian $\hat{H}$ for the structure in Fig.~\ref{FIGschemea} in the polariton basis is given by
\begin{widetext}
\begin{equation}
\label{hamiltonianinsum}
\hat{H}=\displaystyle \sum _{i}\displaystyle \sum_{\mathbf{q}} \hbar \omega_{i} (\mathbf{q}) \hat{c} _{i}^{\dagger} (\mathbf{q}) \hat {c}_{i} (\mathbf{q}) + \frac{g_{0}}{2} \displaystyle \sum _{\substack{i,j \\ k,l}} \displaystyle \sum _{\substack{\mathbf{q}_1, \mathbf{q}_2 \\ \mathbf{q}}} \mathrm{X}_{{i}} (\mathbf{q}_1+\mathbf{q}) \mathrm{X} _{{j}}(\mathbf{q} _2-\mathbf{q}) \mathrm{X}_{{k}} (\mathbf{q}_1)\mathrm{X} _{{l}}(\mathbf{q}_2) \hat{c} _{i}^{\dagger} (\mathbf{q}_1 + \mathbf{q}) \hat{c}_{j} ^{\dagger} (\mathbf{q}_2-\mathbf{q}) \hat{c}_{k} (\mathbf{q}_1) \hat{c}_{l} (\mathbf{q}_2),
\end{equation}
\end{widetext}where $i,j,k,l=\left\{ 1,2,3,4\right\} $ enumerate polariton
branches with corresponding frequencies $\omega _{i}$; $\mathbf{q}$ is an exciton-polariton wave-vector, $\hat{c},\hat{c}%
^{\dagger }$ are the annihilation and creation operators, $\mathrm{X}%
_{i,j,k,l}$ are the Hopfield coefficients defining the exciton fraction in
the polariton state. The nonlinear coupling constant can be approximated by $%
g_{0} \approx \left. 6E_{b} {a_{b}^{3} D}\right/ {d_{\rm QW} V}$, where $E_{b}$ is the
exciton binding energy, $a_{b}$ is the exciton Bohr radius, $D$ is period of
the structure, $d_{\rm QW}$ is the QW width, and $V $ is the
QW area.
Hereafter we restrict ourselves only to LB neglecting the
inter-branch scattering processes. In real space, we will describe LB
polaritons by the field operator $\hat{\Psi}$ defined as
\begin{equation}
\hat{\Psi}(\mathbf{r},t)=\frac{1}{\sqrt{V}}\displaystyle\sum_{\mathbf{q}%
}\hat{c}_{1}(\mathbf{q})e^{i\mathbf{qr}-i\omega _{1}(\mathbf{q})t}.
\label{WFPsi}
\end{equation}%

Next, we use a
mean-field approach to replace the corresponding polariton field operator $%
\hat{\Psi}(\mathbf{r},t)$ by its average value ${\Psi (\mathbf{r},t)=\langle
\hat{\Psi}(\mathbf{r},t)\rangle }$, which characterizes the LB polariton
wave function (WF) associated with the Bragg mirror structure. By using
Eqs.~(\ref{hamiltonianinsum}),~(\ref{WFPsi}) and taking the Fourier
transform we obtain the nonlinear Schr\"{o}dinger equation for the real-space
dynamics of the order parameter ${\Psi (\mathbf{r},t)}$, cf.~\cite{SuM}

\begin{equation}
i\frac{\partial \Psi }{\partial t}=\left[ -\frac{\hbar }{2m_{\Vert }}\Delta
_{\Vert }-\frac{\hbar }{2m_{\perp }}\frac{\partial ^{2}}{\partial z^{2}}%
-i\gamma _{0}+g|\Psi |^{2}\right] \Psi ,  \label{GPEeqnOriginal}
\end{equation}%
where $m_{\Vert }$ and $m_{\perp }$ are the components of the effective mass tensor,
$g = \left. 6E_{b}{a_{b}^{3}} D X_{1}^{4} \right/ \hbar d_{\rm QW} $ is two-body polariton-polariton interaction strength, $X_{1}\approx \left.
\Omega _{P}\right/ 2\Omega _{B}$ is \emph{coordinate independent} Hopfield
coefficient (cf.~\cite{PRA640138092001}), $\Omega _{B}$ is the band gap
half-width, $\Omega _{P}$ is the Rabi-frequency governed by the
exciton-photon coupling strength. We have introduced the dissipation term $%
-i\gamma _{0}$ to account for the radiative decay of polaritons. Deriving
Eq.~(\ref{GPEeqnOriginal}) we assume that $\Omega _{P}$ is much smaller than
$\Omega _{B}$; we also neglect the nonlocal character of polariton-polariton
interaction assuming $X_{1}\equiv X_{1}(0)$. An important peculiarity of our
system is the negative transverse component of the polaritonic effective
mass that tends to ${m_{\perp }=\left. -4\pi ^{2}\hbar \Omega _{B}\right/
\omega _{B}^{2}D^{2}}$, where $\omega _{B}$ is the centre of the second
photonic band gap. Meanwhile, the lateral effective mass $m_{\Vert }\approx 2%
\tilde{\varepsilon}\hbar \omega _{0}/c^{2}$ is positive, $\tilde{\varepsilon}
$ being the average dielectric permittivity of the layered structure.

For the numerical estimations we consider a GaN/AlGaN layered structure with
InGaN QWs, for which the exciton binding energy is approximately $%
45\text{ meV}$, the exciton Bohr radius is ${a_{b}\approx 18}\text{nm}$,
${d_{QW}=10\text{nm}}$
and Rabi frequency
$\Omega _{P}  \approx 2\pi \times 7.25\text{THz}$.
%$\Omega _{P}\approx 30\text{ meV}$.
 We chose the exciton
energy of $2~\text{eV}$ and photonic band gap width of
$0.1\text{eV}$,
$D=125\text{nm}$ and $\tilde{\varepsilon}=$~$5.035$. The polariton decay
rate $\gamma _{0}$ is given by the photonic radiative decay lifetime $\tau
=1/\gamma _{0}$, which is taken to be $0.5~\text{ps}$ that is the typical
value for GaN based microcavities.

\begin{figure}[tbp]
\includegraphics[width= 0.95\columnwidth]{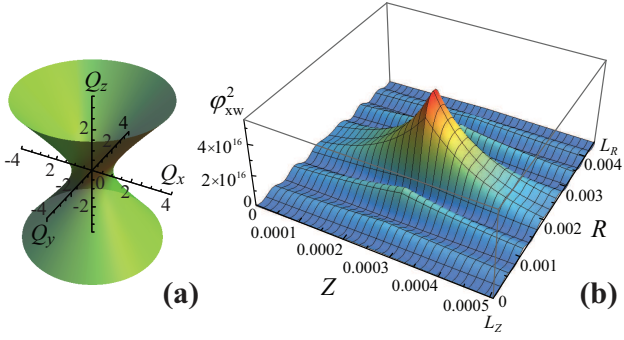}
\caption{(Color online) (a) -- Isofrequency surface for Bragg
exciton-polaritons in the linear dissipationless regime at $g  \simeq 0$, $\protect%
\gamma _0 \simeq 0$. Values of $Q _{X,Y,Z}$ are given in $\protect\sqrt{%
\protect\eta}$ units. The effective masses of polaritons in the structure
are $m _{\|} \approx 3.58 \times 10 ^{-35} \text{kg}$, $m _{\perp} \approx
-3.58 \times 10 ^{-36} \text{kg}$. (b) -- Normalized probability density $%
\protect\varphi _{\mathrm{xw}} ^{2} $ vs spatial $R$ and $Z$ variables; ${(R
- R_{0}) ^{2}} = (X - X_{0}) ^{2} + (Y - Y_{0}) ^{2}$. The parameters are: ${%
\Delta = 6 \times 10 ^{-4} }$, ${Z _{0} = L _{Z} / 2 }$, ${R _{0} = L _{R}/2
= L / \protect\sqrt{2} }$, ${L _{Z} \approx 5.1 \times 10 ^{-4} }$, ${L
\approx 3.2 \times 10 ^{-3} }$, ${\protect\eta \approx 2.6 \times 10^{7} }$,
${N _{\mathrm{in }} \approx 1.98 \times 10 ^{7} }$. Values of the parameters
(except $\Delta$) are taken the same as for Fig.~\protect\ref{FIGpotAndKink}
and are discussed below. }
\label{FIGhypDisp}
\end{figure}

We transform Eq.~(\ref{GPEeqnOriginal}) to the standard Schr\"{o}dinger
equation by introducing new variable $\Psi (\mathbf{r},t)=\phi (\mathbf{r}%
,t)e^{-\gamma _{0}t}$ as
\begin{equation}
i\partial _{t}\phi =\left[ -\frac{\hbar }{2m_{\Vert }}\Delta _{\Vert }+\frac{%
\hbar }{2m}\partial _{zz}+p(t)^{-1}g|\phi |^{2}\right] \phi ,
\label{GPEwithExp}
\end{equation}%
where we denote $m\equiv |m_{\perp }|$. In~(\ref{GPEwithExp}) we suppose
that $e^{-{2\gamma _{0}}t}\approx 1-2\gamma _{0}t\equiv p(t)^{-1}$, cf.~\cite%
{PhysLettA37630572012}. This approach is applicable in the limit of weak
decay: $\gamma _{0}\ll \Omega _{P}$. We focus on the stationary states of LB
polaritons representing the solution of Eq.~(\ref{GPEwithExp}) in the form
\begin{multline}
\varphi (X,Y,Z)=\sqrt{\frac{\kappa ^{2}\kappa _{z}}{p(t)^{3}}}\phi (\mathbf{r%
},t)  \label{varphidef} \\
\times \exp \left[ i\frac{\gamma _{0}m_{\Vert }p(t)}{\hbar }\left(
x^{2}+y^{2}-\frac{m}{m_{\Vert }}z^{2}\right) +iEp(t)t\right] ,
\end{multline}%
where $\kappa = \sqrt{\left.   \hbar V\right/ 2m_{\Vert }g}$,
$\kappa _{z}= \sqrt{\left.  \hbar V \right/ 2mg}$ are characteristic macroscopic scales of
polaritonic system in the structure, $V=L_{x}L_{y}L_{z}$ is volume of the
structure in Fig.~\ref{FIGschemea}, $E$ is energy of the system.
Substituting (\ref{varphidef}) for (\ref{GPEwithExp}) we finally obtain
\begin{equation}
\partial _{ZZ}\varphi -(\partial _{XX} +\partial _{YY})\varphi -\eta
\varphi +G\varphi ^{3}=0,  \label{dimless_f_GPE}
\end{equation}%
where $\eta =\left. EV\right/ g$, $G=\left. V\right/ \kappa ^{2}\kappa _{z}$%
. Polariton WF $\varphi $ obviously obeys a normalization condition
\begin{equation}
%\left. \int_{G}\varphi ^{2}dXdYdZ\right\vert _{t=0}=
\left.
\int_{0}^{L}dX\int_{0}^{L}dY\int_{0}^{L_{Z}}\varphi ^{2}dZ\right\vert
_{t=0}\simeq N_{\mathrm{in}},  \label{NormmCondNInit}
\end{equation}%
where $N_{\mathrm{in}}$ is initial (at ${t=0}$) total number of polaritons,
the dimensionless variables $X=p(t)x/\kappa $, $Y=p(t)y/\kappa $, $%
Z=p(t)z/\kappa _{z}$; $\bar{t}=p(t)t$; $L_{X}=L_{Y}=L$ and $L_{Z}$ are
characteristic dimensionless lengths of the structure.

\emph{Linear regime.} The most interesting features of Eq.~(\ref%
{dimless_f_GPE}) can be elucidated in the linear regime, \emph{i.e.} for the
ideal gas of non-interacting polaritons occurring at $g \simeq 0$ ($\Omega _{P} / \Omega _{B} \rightarrow 0 $) and for the vanishing decay
rate $\gamma _{0}\simeq 0$. In this limit, effective dispersion relation is
obtained from Eq.~(\ref{dimless_f_GPE}) substituting plane wave solution $%
\Psi \propto e^{i\mathbf{QR}}$ : ${\eta =Q_{X}^{2}+Q_{Y}^{2}-Q_{Z}^{2}}$.
The corresponding dispersion surface is shown
in Fig.~\ref{FIGhypDisp}(a). It is clearly seen that a Bragg mirror allows existence of
freely propagating LB polaritons. The specific dispersion of Bragg
polaritons leads to the characteristic HMM divergence of the photonic
density of states~\cite{ContemporaryPhysics542732013}. In the same limit,
Eq.~(\ref{dimless_f_GPE}) supports so called X-wave solution defined as~\cite%
{PRL901704062003}
\begin{equation}
\varphi _{\mathrm{xw}}=\mathbb{C}\text{Re}\left[ v^{-1/2}\exp \left[ -i\sqrt{%
v}\right] \right] ,  \label{XWSolu}
\end{equation}%
where we have redefined $v=\eta \lbrack (\Delta
-i(Z-Z_{0}))^{2}+(X-X_{0})^{2}+(Y-Y_{0})^{2}]$; $\Delta $ is a real-valued
arbitrary coefficient that determines the wave packet
localization, $X_{0}$, $Y_{0}$ and $Z_{0}$ define positions of the center of
the wave packet, $\mathbb{C}$ is normalization constant which can be
estimated using the normalization condition~(\ref{NormmCondNInit}). The
solution~(\ref{XWSolu}) is shown in Fig.~\ref{FIGhypDisp}(b) for the
parameters given above. Physically, the polaritonic X-wave represents a
nondiffracted localized wave packet analogous to diffractionless beams in
optics~\cite{PRL5814991987}.

\begin{figure}[tbp]
\includegraphics[width= 0.9\columnwidth]{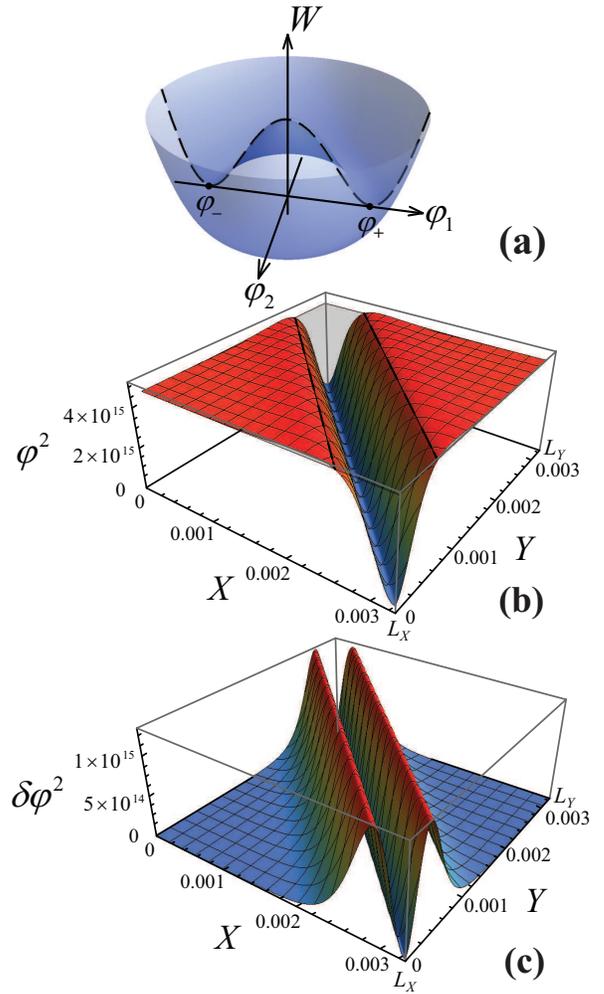}
\caption{(Color online) (a) -- Schematic of Higgs potential $W$ versus $%
\protect\varphi _1$ and $\protect\varphi _2$ variables, (b) -- perturbed
Higgs field (black soliton) $\protect\varphi ^2$ and (c) -- perturbation $%
\protect\delta \protect\varphi ^2$ versus dimensionless spatial coordinates $%
X$ and $Y$. The parameters are: ${X _{0} = Y _{0} = L / 2}$, ${Z = Z _{0}
= \protect\pi / 2 \Omega _{2} }$, ${\protect\mu = 0.2}$, ${L _{x} = L _{y}
= 2\mathrm{\protect\mu}\text{m} }$, ${L _{z} = 1\mathrm{\protect\mu}\text{m%
} }$, ${g / V = 2.49\text{neV} }$, ${E _{1} = 63.8\text{meV} }$. Upper
shadow plane ${\ \protect\phi ^{2} = \protect\eta / G }$ in (b) indicates
vacuum state solution. }
\label{FIGpotAndKink}
\end{figure}

\emph{Polariton Higgs field.} The behavior of our polariton system is
essentially modified in the presence of polariton-polariton scattering,
\emph{i.e.} at $g\neq 0$. Equation~(\ref{dimless_f_GPE}) with the
\textquotedblleft time\textquotedblright\ variable $Z$ represents Ginzburg-Landau-Higgs (GLH)
equation, that is typically discussed in connection with the Universe
properties and bubble evolution \cite{PhysRepC3511978}. In order to study
Eq.~(\ref{dimless_f_GPE}) it is conveninent to represent the Higgs field $%
\varphi $ as a complex scalar field $\varphi =\varphi _{1}+i\varphi _{2}$.
The \textquotedblleft Mexican hat\textquotedblright\ Higgs potential $%
W\equiv W(\varphi _{1},\varphi _{2})$ is shown in Fig.~\ref{FIGpotAndKink}%
(a).
The false vacuum state corresponds to $\varphi =0$ while two real vacuum states are located at
$\varphi _{\pm }=\pm \sqrt{\left. \eta \right/ G} \equiv \pm s$~\cite{RubakovClassicalTheoryofGaugeFields2002}.
These two states correspond to two minima of Higgs field potential. The state $\varphi =0$ is
unstable vs fluctuations, while the states $\varphi _{\pm }$ are stable. The
behavior of a polariton system governed by Eq.~(\ref{dimless_f_GPE}) can be
easily understood if we consider small perturbations $\tilde{\varphi}_{1,2}$
defined by $\varphi _{1}=\varphi _{0}+\tilde{\varphi}_{1}$, $\varphi _{2}=%
\tilde{\varphi}_{2}$ (${\ \tilde{\varphi}_{1,2}\ll \varphi _{0}}$) where $%
\varphi _{0}$ is the ground state solution of Eq.~(\ref{dimless_f_GPE}).

Taking into account the global $U(1)$ symmetry properties of the Lagrangian
for Eq.~(\ref{dimless_f_GPE}) it is possible to conclude that the field $%
\tilde{\varphi}_{1}$ possess a mass whereas the field $\tilde{\varphi}_{2}$
is massless and represents a Nambu-Goldstone boson. Here we focus on $%
\varphi _{1}$ field properties. In order, Eq.~(\ref{dimless_f_GPE})
supports a classical (static) kink or black soliton solution
\begin{equation}
\varphi _{0}(X,Y)=\pm s \tanh \left[ \Theta \right] ,
\label{BlackSlitonSolution}
\end{equation}%
where ${\Theta \equiv \sqrt{\eta }\left. \left( X-X_{0}+Y-Y_{0}\right) \right/ 2}$.
In~(\ref{BlackSlitonSolution}) the parameters $X_{0}$, $Y_{0}$ characterize the position of the envelope minimum. At $X,Y\rightarrow \infty $ the soliton solution in~(\ref{BlackSlitonSolution}) approaches two vacuum states $\varphi _{\pm }$.
Combining~(\ref{NormmCondNInit}) and~(\ref{BlackSlitonSolution}) we obtain a condition: $N_{\mathrm{in}}\simeq L_{Z}(L^{2}\eta -8\ln [\cosh [\tilde{L}]])/G$, that determines the critical number of particles required for a kink formation. Here we assume that ${X_{0}=Y_{0}=L/2}$ and introduce the dimensionless parameter ${\tilde{L}=L\sqrt{\eta }/2}$.

The parameter $s = \sqrt{\left. \eta \right/ G}$ in Eq.~\eqref{BlackSlitonSolution} plays a crucial role in the field theory, cf.~\cite{RubakovClassicalTheoryofGaugeFields2002}.
In the limit $s\gg1$, the soliton can be treated as a classical object.
We consider the oscillon field as a small perturbation for $\varphi $%
-function in $Z$ direction. In order to find it, we represent $\varphi $ in
the form $\varphi =\varphi _{0}+\mu \delta \varphi $ ($\tilde{\varphi}%
_{1}\equiv \mu \delta \varphi $), where the oscillon solution $\delta
\varphi =\delta \varphi (X,Y)\cos \left( \tilde{\Omega}(Z-Z_{0})\right) $
characterizes lateral excitations; we assume that the condition $|\varphi
_{0}|\gg \mu |\delta \varphi |$ is fulfilled. Substituting $\varphi $ and
Eq.~(\ref{BlackSlitonSolution}) in Eq.~(\ref{dimless_f_GPE}) and linearizing
it with respect to $\delta \varphi $, we obtain a Schr\"{o}dinger-like
equation $\hat{F}\delta \varphi (X,Y)=\tilde{\Omega}^{2}\delta \varphi (X,Y)$
for eigenstates ($\delta \varphi $) and eigenvalues ($\tilde{\Omega}$) of
the operator $\hat{F}=-\nabla _{\Vert }^{2}+2\eta -3\eta \text{sech}%
^{2}[\Theta ]$. For $\tilde{\Omega}^{2}=3\eta /2$ the first excited state of
the system is given by:
\begin{equation}
\delta \varphi (X,Y)= s \tanh [\Theta ]\text{sech}[\Theta
].  \label{DeltaPhiPerturb}
\end{equation}

The classical kink state $\varphi _{0}$ becomes perturbed due to low
amplitude oscillations (fluctuations) of\ the Higgs field $\varphi $; the
state being called \textquotedblleft Higgs oscillon\textquotedblright .

In Fig.~\ref{FIGpotAndKink}(b) the perturbed kink $\varphi ^{2}$ as a
function of $X$ and $Y$ at fixed \textquotedblleft time\textquotedblright\
coordinate $Z$ is plotted. At $X\rightarrow \infty $ and $Y\rightarrow
\infty $ the black soliton solution approaches two vacuum states $\varphi
_{\pm }^{2}$ as the shadow plane in Fig.~\ref{FIGpotAndKink}(b) shows.
Taking into account a finite size of the lattice in $X$ and $Y$ directions
and the periodicity of the system in $Z$ direction (with a period of $2\pi /%
\tilde{\Omega}$) we consider the oscillon formation in a 3D box $L_{X}\times
L_{Y}\times L_{Z}$. We write down the condition $\eta =2\pi
^{2}n^{2}/3L_{Z}^{2}$, which is relevant to the normalized energy $\eta
\equiv E_{n}V/g$. The square of a quantized Higgs oscillon amplitude $\delta
\varphi ^{2}$ is plotted in Fig.~\ref{FIGpotAndKink}(c) for the ground state
$(n=1)$.

Let us note that the energy density $J$ of the polariton Higgs field is $J=%
\frac{1}{2}[(\partial _{Z}\varphi )^{2}+(\partial _{X}\varphi
)^{2}+(\partial _{Y}\varphi )^{2}-\eta \varphi ^{2}+\frac{G}{2}\varphi ^{4}]$%
, while the energy density $J_{0}$ of the kink is $J_{0}=\frac{1}{2}\left[
(\partial _{X}\varphi _{0})^{2}+(\partial _{Y}\varphi _{0})^{2}-\eta \varphi
_{0}^{2}+\frac{G}{2}\varphi _{0}^{4}\right] $. Since ${|\varphi _{0}|\gg \mu
|\delta \varphi |}$, $J$ approaches $J_{0}$. Integrating $J_{0}$ over the
space coordinates $X$, $Y$ we obtain the energy density in $Z$ direction as
\begin{equation}
E_{0,Z}=\frac{\eta }{3G}\left( 2-3\tilde{L}^{2}+8\ln \left[ \cosh [\tilde{L}]%
\right] -2\text{sech}^{2}[\tilde{L}]\right) .  \label{energydensInZdir}
\end{equation}

The energy density of \textquotedblleft vacuum\textquotedblright\ states $%
\varphi _{\pm }$ is $J_{\pm }=\frac{1}{2}\left[ -\eta \varphi _{\pm }^{2}+%
\frac{G}{2}\varphi _{\pm }^{4}\right] $. Hence we can write down energy
density in $Z$ direction as $E_{\pm ,Z}=-\tilde{L}^{2}\eta /G$.
%Finally, taking into account Eq.~(\ref{energydensInZdir}) we can find the so-called mass of the kink $M = E_{0,Z} - E _{\pm,Z}$ as
%\begin{equation}
%\label{KinkMass}
%M = \frac{\eta}{3G} \left( 2+ 8 \ln \left[  \cosh[\tilde{L}] \right] - 2 \text{sech} ^2 [\tilde{L}]  \right) .
%\end{equation}
Taking into account Eq.~(\ref{energydensInZdir}) we can introduce the so-called ``mass'' of the kink $M \sim E_{0,Z}-E_{\pm ,Z}$.
Physically static dark soliton behaves as a relativistic particle with energy $E=Mc^{2}$ at rest, where $c$ is speed of light, cf.~\cite{PhysRepC3511978}.
Note that the ``mass'' of the soliton~$M$ is dependent on the size of the lattice structure.

\emph{Topological Schr\"{o}dinger cat states.} Now let us discuss the Higgs
field properties beyond the mean field theory. The quantum tunneling between
states $\varphi _{\pm }$ representing two minima of the Higgs potential,
Fig.~\ref{FIGpotAndKink}(a) is responsible for creation of field bubbles in
the gauge field theory~\cite{RubakovClassicalTheoryofGaugeFields2002}.
The tunneling leads to formation of the Schr\"{o}dinger cat states
(superposition states):
\begin{equation}
\left\vert \psi _{\pm }\right\rangle =\frac{1}{\sqrt{2(1+e^{- 2 s^{2}})}} \left( \left\vert \phi _{+}\right\rangle \pm \left\vert \phi
_{-}\right\rangle \right) ,  \label{psistate}
\end{equation}%
where $\left\vert \phi _{+}\right\rangle $ and $\left\vert \phi
_{-}\right\rangle $ are macroscopically distinguishable Glauber's coherent
states associated with fields $\phi _{+}$ and $\phi _{-}$, respectively.
The ``size of the cat'' can be estimated via the overlap integral of the states $| \phi _{\pm}\rangle$ as $ \zeta = 1/ \langle \phi _{+} | \phi _{-} \rangle = e ^{2 s ^{2}}$ \cite{PhysRevA5712081998}.
The parameter $\zeta$  becomes larger in the limit $s \gg 1$  which is indeed experimentally achievable in realistic structures~\eqref{BlackSlitonSolution}.
Notably, the properties of states $\left\vert \psi _{\pm }\right\rangle $
are highly non-classical, see e.g.~\cite{PRL57131986}. In particular, due to
the interference, a fringe pattern occurs between Gaussian bells
representing states $\left\vert \phi _{\pm }\right\rangle $ in the Wigner
function approach. The negativity of this function that is inherent to the
states~(\ref{psistate}) is responsible for that. While the states~(\ref%
{psistate}) involve a macroscopically large number of particles they can be
used for generation of macroscopic entangled states (cf.~\cite{PRA4568111992}%
) of exciton-polaritons in Bragg-superlattices. The computational qubit
states $\left\vert 0\right\rangle $ and $\left\vert 1\right\rangle $ can be
associated with mutually orthogonal states~(\ref{psistate}) as $\left\vert
0\right\rangle =\left\vert \psi _{+}\right\rangle $ and $\left\vert
1\right\rangle =\left\vert \psi _{-}\right\rangle $, cf.~\cite%
{PRA610423092000}. Alternatively, if the parameter $e^{-2s^{2}}$ in~(\ref%
{psistate}) vanishes rapidly, the topological states $\left\vert \phi _{+}\right\rangle $ and $\left\vert \phi _{-}\right\rangle $ for the quantum
Higgs field itself may be considered as a computational qubit states $%
\left\vert 0\right\rangle $ and $\left\vert 1\right\rangle $~\cite%
{PRA680423192003}. In this case qubit operations presume implementation of
linear circuit networks and conditional photon detection~\cite%
{Nature409462001}. Amazingly, such circuits can be designed using well
developed semiconductor technologies, ~\cite%
{NatCommun417782013,PRL1121964032014}.

In conclusion, we propose realisation of quantum HMMs in a periodic planar
semiconductor Bragg mirror with embedded QWs. We demonstrate
mapping of the polaritonic Gross-Pitaevskii equation onto a nonlinear
Ginzburg-Landau-Higgs equation, which exhibits physically non-trivial
features. In the liner case, i. e. for non-interacting LB polaritons
we obtain a polariton X-wave solution that is reminiscent of a
non-diffractive (spatially localized) matter wave packet. We predict
formation of kink-shaped states for weakly interacting polaritons. Small
amplitude oscillations (oscillons) occur in a perturbed polariton Higgs
field due to fluctuations. Going beyond the mean field theory we obtain a
Schr\"{o}dinger cat state as a macroscopic  superposition of two vacuum states $\varphi
_{\pm }$. Polaritonic nonlinear HMMs have a high potentiality for
simulation of fundamental cosmological processes.

\begin{acknowledgments}
This work was supported by RFBR Grants No.~14-02-31443, No.~14-02-92604, No.~14-32-50420, No.~14-02-97503,
No.~15-52-52001, No.~15-59-30406,
by the Russian Ministry of Education and Science state tasks No. 2014/13, 16.440.2014/K,
by President grant for leader scientific school No. 89.2014.2 and EU project PIRSES-GA-2013-612600 LIMACONA.
A.P.A. acknowledges support from ``Dynasty'' Foundation.
\end{acknowledgments}

\end{document}